\documentclass{iau} 
\usepackage{graphicx,wrapfig,url,amsmath,wasysym,xcolor}

\newcommand{\urlwofont}[1]{\urlstyle{same}\url{#1}}

\newcommand{\SigmaSFR}{\Sigma_{\mathrm{SFR}}}
\newcommand{\epsff}{\epsilon_{\mathrm{ff}}}
\newcommand\Msun{\text{M}_{\astrosun}} 

\title[Origin of globular clusters] 
{Formation and evolution of globular clusters in cosmological simulations}

\author[Hui Li \& Oleg Gnedin]   
{Hui Li$^1$
 \and Oleg Gnedin$^2$}

\affiliation{$^1$Department of Physics, Kavli Institute for Astrophysics and Space Research,\\MIT, Cambridge, MA 02139, USA \\ email: {\tt hliastro@mit.edu} \\[\affilskip]
$^2$Department of Astronomy, University of Michigan,\\Ann Arbor, MI 48109, USA \\email: {\tt ognedin@umich.edu}}

\pubyear{2019}
\volume{351}  
\jname{Star Clusters: From the Milky Way to the Early Universe}
\editors{A. Bragaglia, M.B. Davies, A. Sills \& E. Vesperini, eds.}
\begin{document}

\maketitle

\begin{abstract}
In a series of three papers, we introduced a novel cluster formation model that describes the formation, growth, and disruption of star clusters in high-resolution cosmological simulations. We tested this model on a Milky Way-sized galaxy and found that various properties of young massive clusters, such as the mass function and formation efficiency, are consistent with observations in the local universe. Interestingly, most massive clusters -- globular cluster candidates -- are preferentially formed during major merger events. We follow the dynamical evolution of clusters in the galactic tidal field. Due to tidal disruption, the cluster mass function evolves from initial power law to a peaked shape. The surviving clusters at $z=0$ show a broad range of metallicity [Fe/H] from -3 to -0.5. A robust prediction of the model is the age--metallicity relation, in which metal-rich clusters are systematically younger than metal-poor clusters by up to 3 Gyr.

\end{abstract}

\firstsection 
\section{Introduction}

Globular clusters (GCs) are commonly found in galaxies with stellar mass larger than $10^9\Msun$ (Brodie \& Strader 2006). The tight correlation between the total mass of the GC system and host galaxy halo mass and the broad range of GC metallicity distribution suggest that GCs trace well the mass assembly and metal enrichment history of their host galaxies. Therefore, the co-evolution of GCs and galaxies can only be understood within the framework of hierarchical structure formation.

Recent endeavors have been made to incorporate galaxy formation physics with the formation and evolution of GCs in cosmological contexts. The first type of these models is so-called ``semi-analytical'' model, where GCs are painted on top of the preexisting cosmological simulations, either N-body (e.g. Muratov \& Gnedin 2010, Li \& Gnedin 2014, Choksi, Gnedin, \& Li 2018, El-Badry et al. 2019) or hydrodynamical (e.g. Pfeffer et al. 2018, Ramos-Almendares et al. 2019). The advantage of this approach is that it is relatively computationally inexpensive so that one can study the properties of GC populations in a large number of galaxies over a wide mass range. However, these models usually rely on specific assumptions of the formation physics of massive clusters that cannot be resolved by the simulations themselves. Another approach is to perform cosmological hydrodynamical simulations with extremely high spatial/mass resolutions so that massive star clusters can be roughly resolved (but not dynamically) by an ensemble of star particles (e.g. Kim et al. 2018, Ma et al. 2019, Lahen et al. 2019). These simulations have proved to be powerful in pinpointing the formation sites of a few massive clusters in highly turbulent gas-rich disks. Due to the extreme computational cost, however, they are limited to high redshifts and cannot follow the long-term dynamical evolution of the model clusters.

Here we developed a novel star formation prescription in cosmological simulations by considering star clusters as a unit of star formation. This new approach allows us to model the physical process of star clusters formation from the first principle and, at the same time, track the subsequent dynamical evolution of the whole cluster population across cosmic time.

\begin{figure}[b]
\begin{center}
 \includegraphics[width=3.in]{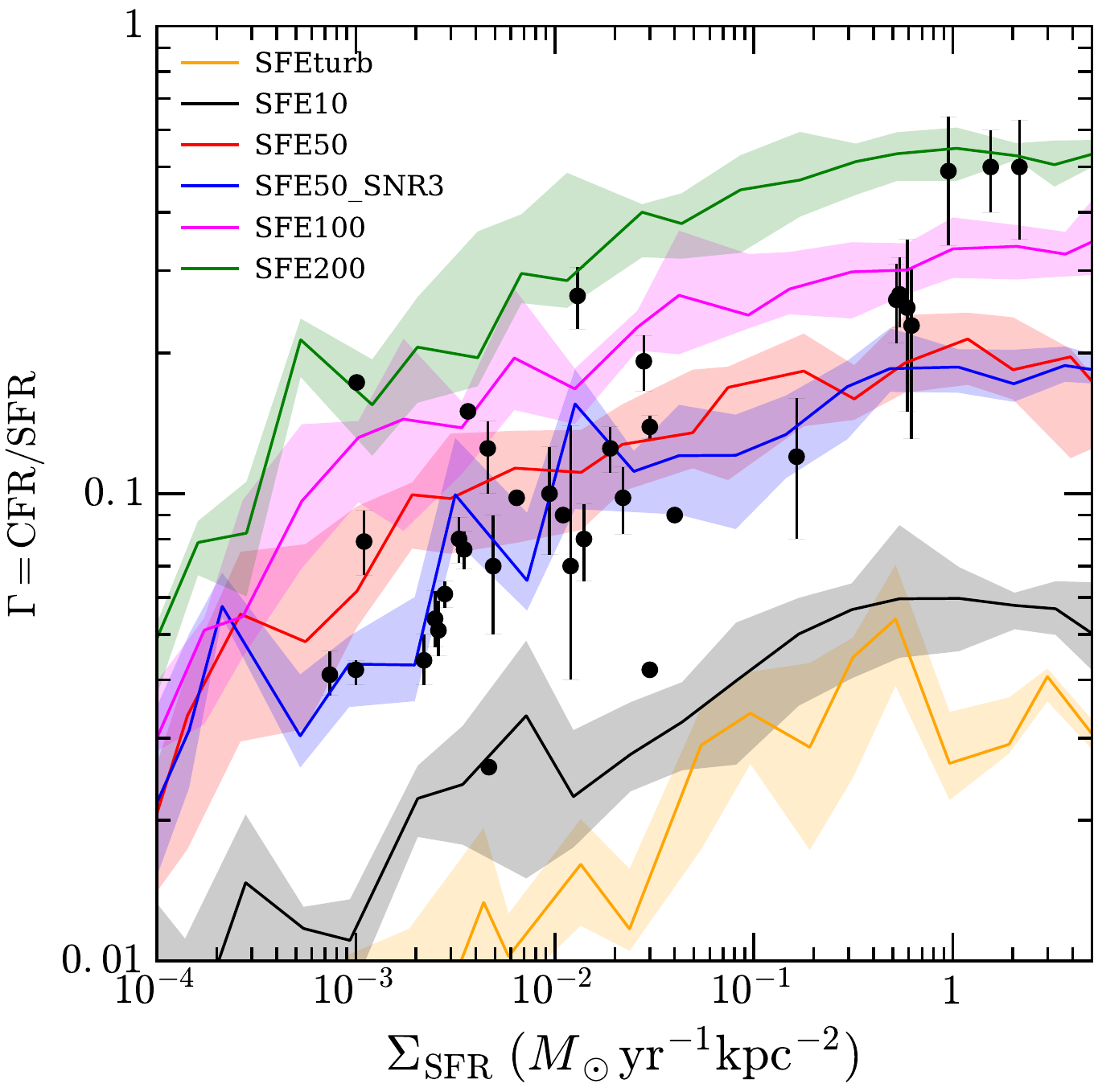} 
 \caption{Fraction of clustered star formation as a function of SFR surface density. Solid lines and shaded areas show the median and 25\%--75\% interquartile range of the distribution of $\Gamma$ for a given $\Sigma_{\rm SFR}$ bin in simulations with different $\epsff$. The observed values (symbols with error bars) are from a compilation of both galaxy-wide and spatially resolved measurements of cluster samples in nearby galaxies.}
  \label{fig:gamma}
\end{center}
\end{figure}

\section{Star cluster as a unit of star formation in cosmological hydrodynamic simulations}

We implemented the model in the mesh-based Adaptive Refinement Tree code (Li et al. 2017, 2018).
In this model, cluster particles are first seeded at the density peaks of the galactic disk, and then grow in mass continuously via gas accretion from its 27 neighboring cells at a rate determined by local gas properties with an efficiency $\epsff$.
The mass growth is resolved with high time resolution and is terminated by its own energy and momentum feedback; thus, the final cluster mass is determined self-consistently and represent the mass of star clusters emerged from their natal gas clouds. We tested this model in cosmological simulations of a Milky Way-sized galaxy with 5 parsec spatial resolution, including many physical ingredients such as radiative transfer, non-equilibrium cooling/heating/chemistry, and stellar feedback from stellar winds, ionizing radiation, and supernovae.

\section{Environment-dependent star cluster formation}

We first examined the properties of young massive clusters (YMCs) formed in the simulations and found that the simulations reproduced various properties of both the host galaxy and YMCs.
For example, the cluster initial mass function (CIMF) develops a power-law distribution with an exponential cut-off at the high mass end. The power-law slope of around --2 is similar to the observed value in nearby galaxies. We also found that the cut-off mass increases with the star formation rate of the host galaxy, suggesting an environment-dependent cluster formation scenario, ruling out the idea that cluster mass is randomly sampled from a universal CIMF (Li et al. 2017).

\begin{figure}[b]
\begin{center}
 \includegraphics[width=2.5in]{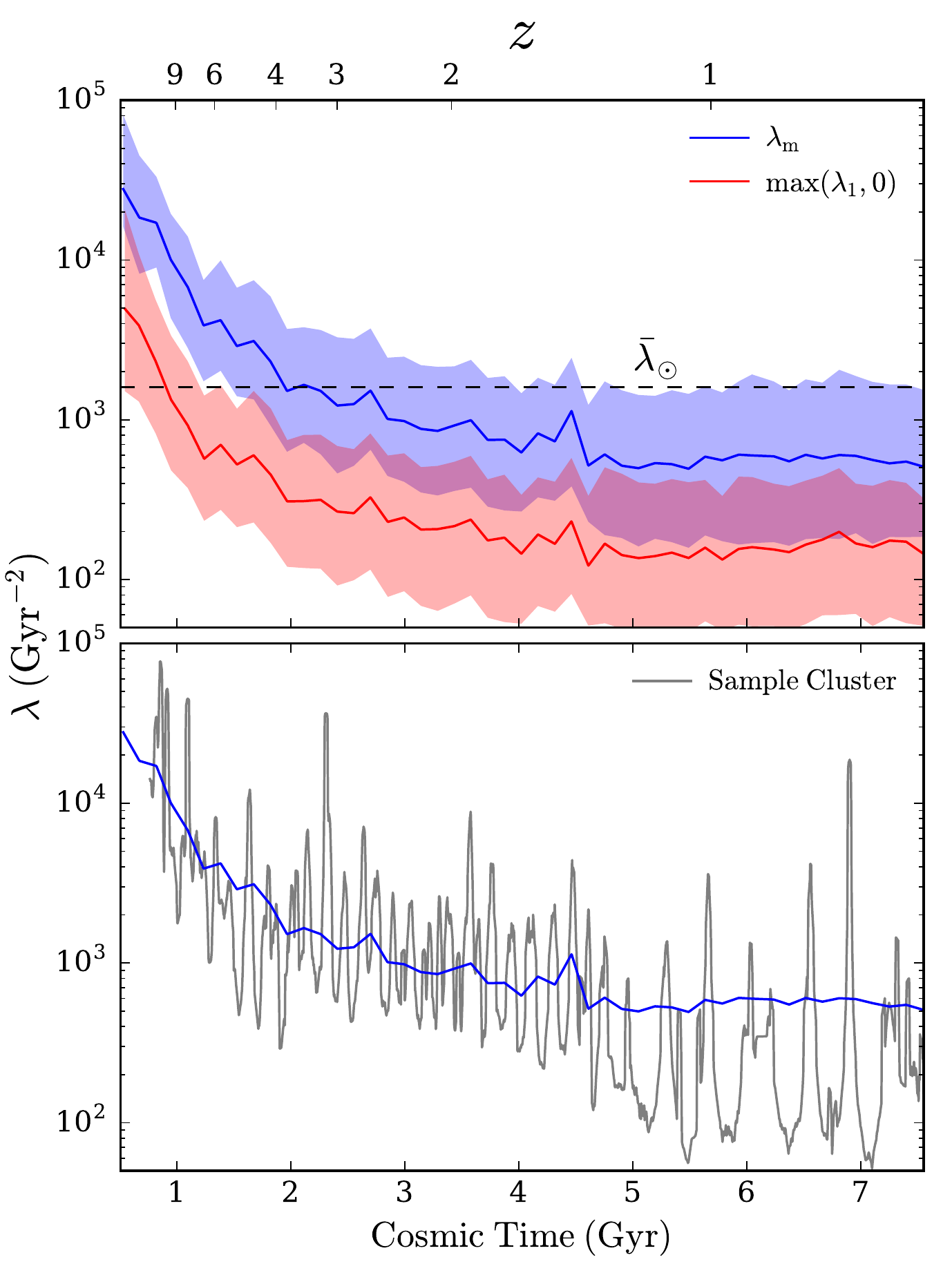} 
 \caption{Upper: Time evolution of the eigenvalues of the tidal tensor, for clusters more massive than $10^5\Msun$ in SFE200 run. Blue and red lines show the median values of $\lambda_m$ and of positive values of $\lambda_1$, respectively, as a function of cluster age, while the shaded areas show the 25\%--75\% interquartile range. Lower: Evolution of $\lambda_m$ with high time resolution for the same sample cluster (black line).}
  \label{fig:tides}
\end{center}
\end{figure}

The simulations revealed a positive correlation between the star formation rate surface density $\SigmaSFR$ of the galactic disk and the cluster formation efficiency ($\Gamma$, defined as the mass fraction of young stars contained in bound clusters). 
This positive correlation is consistent with the observed trend from nearby galaxies (Adamo et al. 2015).
It is also clear that the normalization of $\Gamma$ increases with $\epsff$. 
This trend makes $\Gamma$ an excellent observable and limits the choice of $\epsff$ to 0.5--1. This constraint on $\epsff$ is from the scale of star-forming regions and is independent to the ones based on overall galactic properties.
 
\section{Major mergers as the factory of GC progenitors}

It has long been suggested that major mergers between galaxies can trigger the formation of most massive star clusters that potentially survive the cosmic time and become GCs (Ashman \& Zepf 1992). Here, we investigate the effects of major mergers on CIMF. We found that major mergers change the shape of the CIMF in two ways: the power-law slope is shallower and the cutoff mass is higher in major mergers than that in quiescent. The combined effect leads to an enhancement of the formation of most massive clusters during major mergers.

To systematically evaluate the effects of major mergers across cosmic time, we identified all massive clusters formed during the three major mergers in the simulations. Roughly 75\% of all massive clusters with $M> 2\times10^5\Msun$ are formed during the three major mergers experienced by the host galaxy. A large fraction of these massive clusters survive and become GCs at present. In contrast, the formation of clusters with $M < 2\times10^5\Msun$ does not correlate with major merger events but follows instead the overall star formation history of the host galaxy. This finding supports the merger-induced GC formation scenario, an assumption that has been adopted to build our previous semi-analytical models (Muratov \& Gnedin 2010; Li \& Gnedin 2014; Choksi et al. 2018; Choksi \& Gnedin 2019).

\section{Tidal disruption of star clusters across cosmic time}

\begin{figure}[b]
\begin{center}
 \includegraphics[width=0.477\linewidth]{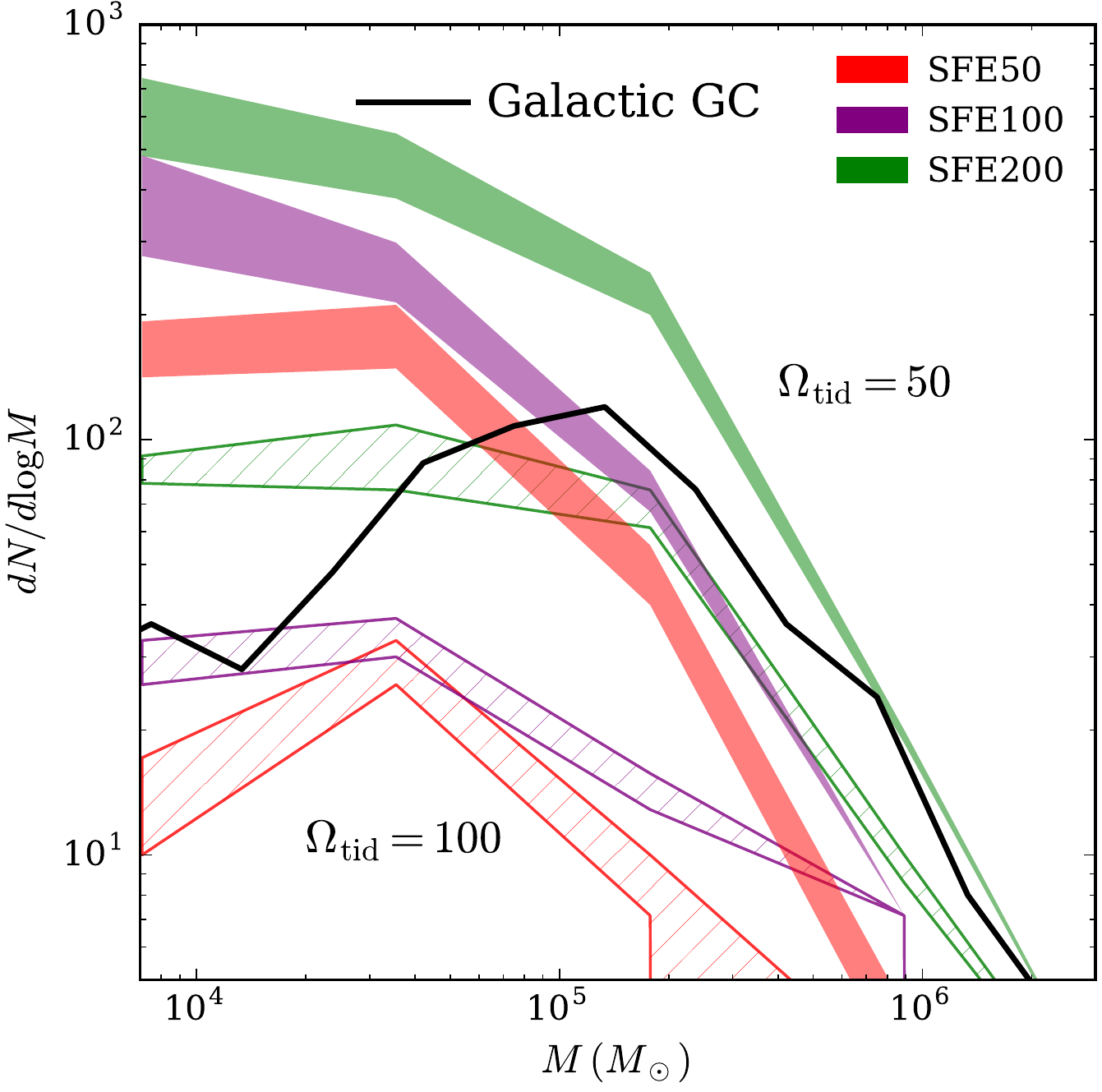}
 \includegraphics[width=0.49\linewidth]{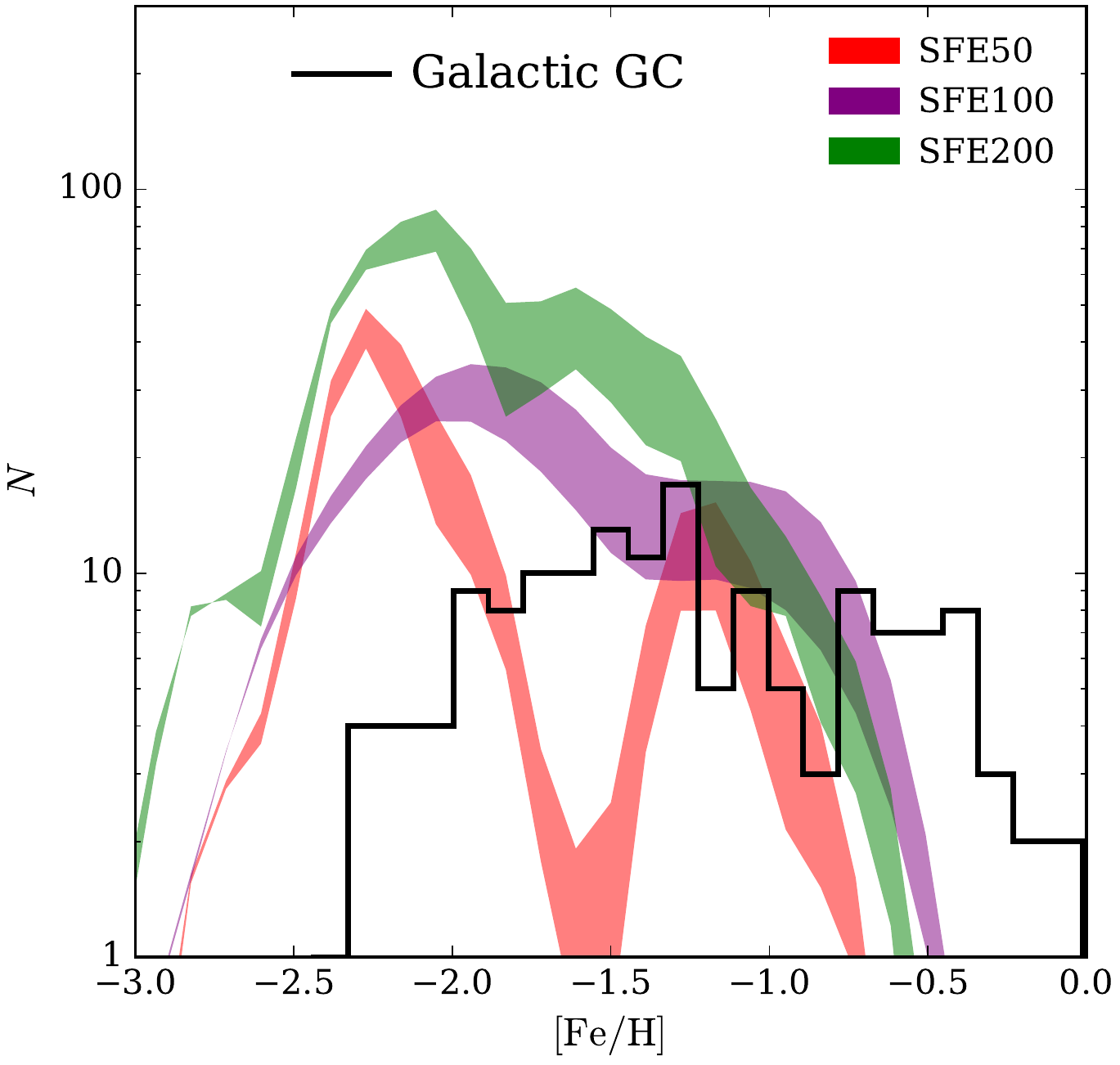}
 \caption{Left: Mass function of surviving clusters at $z=0$ for runs SFE50 (red), SFE100 (purple), SFE200 (green). Contours for each run represent the range of the mass function using two ways of estimating the strength of the tidal field ($\lambda_m$ and $\lambda_1$). Mass function of the Galactic GCs is plotted as black line. Right: Metallicity distribution of surviving star clusters (with remaining mass higher than $10^4\Msun$) in the main galaxy at $z=0$. Metallicity distribution of Galactic GCs is shown as black line for reference.}
  \label{fig:GC-properties}
\end{center}
\end{figure}

The realistic modeling of the formation of YMCs in our simulations provides us an ideal starting point to investigate their subsequent dynamical evolution in order to better connect the YMCs formed at high redshifts and GCs at present. In Li \& Gnedin (2019), we introduced a method to calculate the tidal field along the orbit of model clusters by evaluating the time-varying tidal tensors as the second spatial derivative of the gravitational potential: $-\partial^2\Phi/\partial r_\alpha \partial r_\beta$. We estimated the tidal disruption rate of model clusters based on a numerical fit from of N-body simulations for clusters of a given mass and external tidal field. This was the first time that self-consistent star cluster formation and disruption processes were calculated simultaneously during the runtime of cosmological simulations.

We found that YMCs experience very strong tides during the first Gyr when the clusters located within the dense gaseous disk of the host galaxy. The maximum eigenvalues of the tensor, $\lambda_{\rm m}$, during this period are on average one or two orders of magnitude higher than those around the solar neighborhood. It later drops dramatically when the cluster orbits are lifted to larger radii by interactions with dense gas clumps and major mergers (see Fig.~\ref{fig:tides}).
This tidal evolution history suggests that YMCs are most vulnerable during the first Gyr after their formation. We emphasize that, in order to obtain the strong tides in the early time, the simulations require parsec-scale spatial resolution so that the cold dense gas in the gaseous disk is well resolved. Simulations with coarser resolution inevitably underestimate the tidal strength and, in turn, the cluster disruption rate.

\section{GC properties at $z=0$}

\begin{figure}[b]
\begin{center}
 \includegraphics[width=3.4in]{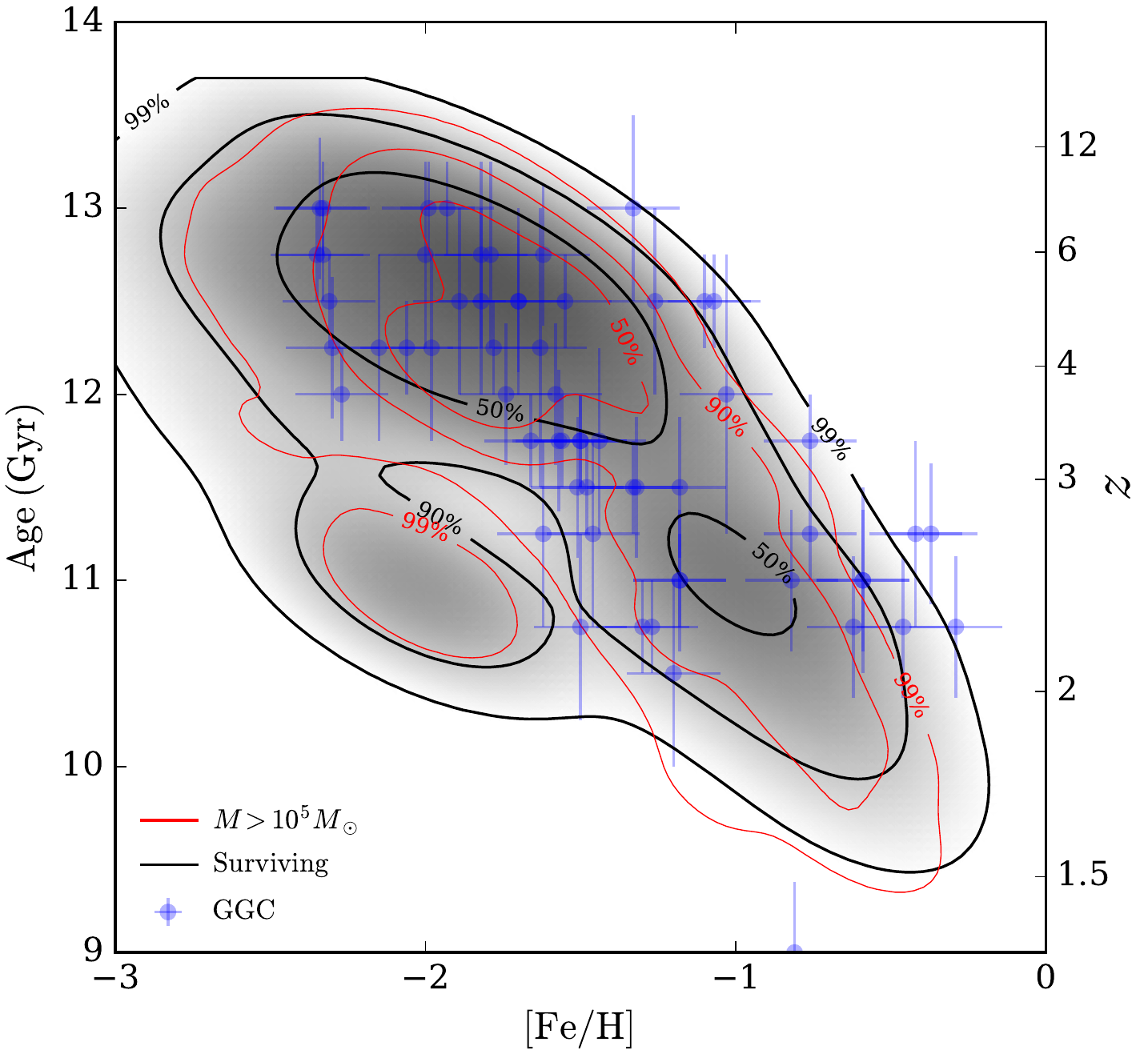}
 \caption{Age–metallicity distribution of model clusters in run SFE100 with initial mass higher than $10^5\Msun$ (red) and of surviving clusters with remaining mass higher than $10^4\Msun$ at $z=0$ (black). The percentage levels shown at each contour line represent the fraction of the number of model clusters enclosed within the corresponding contour. The observed age and metallicity of the Galactic GCs from Leaman, VandenBerg \& Mendel (2013) are overplotted for comparison (blue circles with errorbars).}
  \label{fig:age-metal}
\end{center}
\end{figure}
We investigated the evolution of the mass and metallicity distributions of surviving clusters. Due to tidal disruption, the cluster mass function evolves from an initial truncated power law to a peaked distribution at $z=1.5$. The mass function of GCs at $z=0$ is similar to that of the observed Galactic GCs, although the simulations over-produce low-mass clusters with $M<10^5\Msun$, see the left panel of Fig.~\ref{fig:GC-properties}. This discrepancy may be due to a lack of mass-loss implementation during tidal shocks when clusters cross the gaseous disks, which will be improved in a future work. Another uncertainty of the mass function of bound clusters comes from the physical model of initial bound fraction $f_{\rm i}$ of clusters when they emerge from their natal clouds. It is generally believed that $f_{\rm i}$ depends on the integrated star formation efficiency of the star-forming regions $\epsilon_{\rm int}$, but quantitatively the relationship between the two is not well-understood. In Li et al. (2019), we performed a series of hydrodynamical simulations of individual giant molecular clouds and explored the $f_{\rm i}-\epsilon_{\rm int}$ relation over a wide range of cloud mass and size. We constructed an analytical expression on $f_{\rm i}-\epsilon_{\rm int}$ relation that can be used to improve the sub-grid cluster formation model in cosmological simulations.

It is clear in the simulations that tidal disruption destroys majority of clusters, but does not change the shape of the metallicity distribution much. Both the overall cluster population and the surviving clusters at $z=0$ show a broad range of metallicity from $\rm [Fe/H]=-3$ to $-0.5$ (right panel of Fig.~\ref{fig:GC-properties}), which is similar to the observations. A closer look at the distribution suggests that the model produces more metal-poor clusters and not enough very metal-rich cluster to fully match the observations. One way to push the model metallicity distribution to higher values is to form clusters at later times in more massive, more metal-rich galaxies. However, in order to reach $\rm [Fe/H] \approx 0$, GCs would need to form at $z<1$. The other way is to raise the galaxy-wide metallicity. However, the galaxy stellar mass-metallicity relation obtained from our simulations is a natural consequence of metal enrichment via large-scale galactic outflows due to efficient stellar feedback. Reducing the strength of feedback increases the metallicity by suppressing outflows, but at the cost of higher star formation rate that is inconsistent with the abundance matching result.
It should noted that the current simulations are from only one realization of the Milky Way-sized galaxies. New ``Milky Way'' zoom-in simulations are underway now and the variation of GC properties due to different mass assembly histories will be explored in future work (G.~Brown et al. in prep.).

The surviving GCs exhibit a clear age-metallicity relation, where metal-rich clusters are systematically younger than metal-poor clusters by up to 3 Gyr, see Fig.~\ref{fig:age-metal}. The overall age-metallicity distribution of model clusters is consistent with observations, suggesting that YMCs formed at high redshift are promising GC candidates. This relation is also consistent with the predictions of the semi-analytical model of GC formation. It is a robust prediction emerged naturally from the hierarchical structure formation of galaxies.


\begin{thebibliography}{}
\bibitem[\protect\citeauthoryear{Adamo, Kruijssen, Bastian, Silva-Villa \& Ryon}{2015}]{2015MNRAS.452..246A} Adamo A., Kruijssen J.~M.~D., Bastian N., Silva-Villa E., Ryon J., 2015, MNRAS, 452, 246

\bibitem[\protect\citeauthoryear{Ashman \& Zepf}{1992}]{1992ApJ...384...50A} Ashman K.~M., Zepf S.~E., 1992, ApJ, 384, 50

\bibitem[\protect\citeauthoryear{Brodie \& Strader}{2006}]{2006ARA&A..44..193B} Brodie J.~P., Strader J., 2006, ARA\&A, 44, 193

\bibitem[\protect\citeauthoryear{Choksi, Gnedin \& Li}{2018}]{2018MNRAS.480.2343C} Choksi N., Gnedin O.~Y., Li H., 2018, MNRAS, 480, 2343

\bibitem[\protect\citeauthoryear{Choksi \& Gnedin}{2019}]{2019MNRAS.486..331C} Choksi N., Gnedin O.~Y., 2019, MNRAS, 486, 331

\bibitem[\protect\citeauthoryear{El-Badry, Quataert, Weisz, Choksi \& Boylan-Kolchin}{2019}]{2019MNRAS.482.4528E} El-Badry K., Quataert E., Weisz D.~R., Choksi N., Boylan-Kolchin M., 2019, MNRAS, 482, 4528

\bibitem[\protect\citeauthoryear{Kim, et al.}{2018}]{2018MNRAS.474.4232K} Kim J., et al., 2018, MNRAS, 474, 4232

\bibitem[\protect\citeauthoryear{Lah{\'e}n, Naab, Johansson, Elmegreen, Hu \& Walch}{2019}]{2019ApJ...879L..18L} Lah{\'e}n N., Naab T., Johansson P.~H., Elmegreen B., Hu C.-Y., Walch S., 2019, ApJL, 879, L18

\bibitem[\protect\citeauthoryear{Leaman, VandenBerg \& Mendel}{2013}]{2013MNRAS.436..122L} Leaman R., VandenBerg D.~A., Mendel J.~T., 2013, MNRAS, 436, 122

\bibitem[\protect\citeauthoryear{Li \& Gnedin}{2014}]{2014ApJ...796...10L} Li H., Gnedin O.~Y., 2014, ApJ, 796, 10
\bibitem[\protect\citeauthoryear{Li, Gnedin, Gnedin, Meng, Semenov \& Kravtsov}{2017}]{2017ApJ...834...69L} Li H., Gnedin O.~Y., Gnedin N.~Y., Meng X., Semenov V.~A., Kravtsov A.~V., 2017, ApJ, 834, 69

\bibitem[\protect\citeauthoryear{Li, Gnedin \& Gnedin}{2018}]{2018ApJ...861..107L} Li H., Gnedin O.~Y., Gnedin N.~Y., 2018, ApJ, 861, 107

\bibitem[\protect\citeauthoryear{Li \& Gnedin}{2019}]{2019MNRAS.486.4030L} Li H., Gnedin O.~Y., 2019, MNRAS, 486, 4030

\bibitem[\protect\citeauthoryear{Li, Vogelsberger, Marinacci \& Gnedin}{2019}]{2019MNRAS.487..364L} Li H., Vogelsberger M., Marinacci F., Gnedin O.~Y., 2019, MNRAS, 487, 364

\bibitem[\protect\citeauthoryear{Ma, et al.}{2019}]{2019arXiv190611261M} Ma X., et al., 2019, arXiv, arXiv:1906.11261

\bibitem[\protect\citeauthoryear{Muratov \& Gnedin}{2010}]{2010ApJ...718.1266M} Muratov A.~L., Gnedin O.~Y., 2010, ApJ, 718, 1266

\bibitem[\protect\citeauthoryear{Pfeffer, Kruijssen, Crain \& Bastian}{2018}]{2018MNRAS.475.4309P} Pfeffer J., Kruijssen J.~M.~D., Crain R.~A., Bastian N., 2018, MNRAS, 475, 4309

\bibitem[\protect\citeauthoryear{Ramos-Almendares, Sales, Abadi, Doppel, Muriel \& Peng}{2019}]{2019arXiv190611921R} Ramos-Almendares F., Sales L.~V., Abadi M.~G., Doppel J.~E., Muriel H., Peng E.~W., 2019, arXiv, arXiv:1906.11921

\end{thebibliography}
\end{document}